\documentclass[twocolumn]{aastex62}

\newcommand{\Alf}{Alfv\'{e}n}

\usepackage{multirow}
\hypersetup{citecolor=blue,urlcolor=blue}
\usepackage{upgreek}

\usepackage{amssymb}% http://ctan.org/pkg/amssymb
\usepackage{pifont}% http://ctan.org/pkg/pifont
\shorttitle{3D magnetic field structure in Ursa Major}
\shortauthors{A. Tritsis et al.}

\begin{document}

%===========================================================================
\title{Magnetic field tomography in two clouds towards Ursa Major using H\textsc{I} fibers}
%The dispersion relation of molecular cloud striations: Measurements of temporal variations, total magnetic field strength and projection angle
%===========================================================================

\correspondingauthor{Aris Tritsis}
\email{Aris.Tritsis@anu.edu.au}

\author[0000-0003-4987-7754]{Aris Tritsis}
\affil{Research School of Astronomy and Astrophysics, Australian National University, Canberra, ACT 2611, Australia}

\author{Christoph Federrath}
\affil{Research School of Astronomy and Astrophysics, Australian National University, Canberra, ACT 2611, Australia}

\author{Vasiliki Pavlidou}
\affil{Department of Physics and  Institute for Theoretical and C
omputational Physics, University of Crete, 70013, Heraklion, Greece}
\affil{Foundation for Research and Technology Hellas, IESL \& Institute of Astrophysics, Voutes, 70013, Heraklion, Greece}

\begin{abstract}

The atomic interstellar medium (ISM) is observed to be full of linear structures, referred to as ``fibers". Fibers exhibit similar properties to linear structures found in molecular clouds, termed striations. Suggestive of a similar formation mechanism, both striations and fibers appear to be ordered, quasi-periodic and well-aligned with the magnetic field. The prevailing formation mechanism for striations involves the excitation of fast magnetosonic waves. Based on this theoretical model, and through a combination of velocity centroids and column density maps, Tritsis et al. (2018) developed a method for estimating the plane-of-sky (POS) magnetic field from molecular cloud striations. We apply this method in two H\textsc{I} clouds with fibers along the same line-of-sight (LOS) towards the ultra-high-energy cosmic-ray (UHECR) hotspot, at the boundaries of Ursa Major. For the cloud located closer to Earth, where Zeeman observations from the literature were also available, we find general agreement in the distributions of the LOS and POS components of the magnetic field. We find relatively large values for the total magnetic field (ranging from $\sim$$\rm{10}$ to $\sim$$\rm{20} ~\rm{\upmu G}$) and an average projection angle with respect to the LOS of $\sim$ 50$^\circ$. For the cloud located further away, we find a large value for the POS component of the magnetic field of $15^{+8}_{-3}~\rm{\upmu G}$. We discuss the potential of our new magnetic-field tomography method for large-scale application. We consider the implications of our findings on the accuracy of current reconstructions of the Galactic magnetic field and on the propagation of UHECR through the ISM. 
\end{abstract}

%=-=-=-=-=-=-=-=-=-=-=-=-=-=-=-=-=-=-=-=-=-=-=-=-=-=-=-=-=-=-=-=-=-=-=%
%=-=-=-=-=-=-=-=-=-=-=-=-=-=-=-=-=-=-=-=-=-=-=-=-=-=-=-=-=-=-=-=-=-=-=%
\section{Introduction}\label{intro}

Large Galactic H\textsc{I} surveys such as the Galactic Arecibo L-Band Feed Array H\textsc{I} (GALFA-H\textsc{I}) (Peek et al. 2011; Peek et al. 2018), Parkes Galactic All Sky Survey (GASS) (McClure-Griffiths et al. 2009; Kalberla et al. 2010; Kalberla \& Haud 2015) and Effelsberg--Bonn H\textsc{I} Survey (EBHIS) (Kerp et al. 2011; Winkel et al. 2016) have revealed the ubiquitous presence of parallel, elongated, quasi-periodic structures in H\textsc{I} clouds. These structures, referred to as ``fibers" in the literature, are observed to be well aligned with the plane-of-sky (POS) magnetic field as this is probed by polarization measurements (McClure-Griffiths et al. 2006; Clark et al. 2014; Clark et al. 2015; Planck Collaboration et al. 2016a; Clark 2018; Jeli{\'c} et al. 2018).

Fibers bear important similarities to structures found in the outskirts of molecular clouds, dubbed ``striations". Striations are low-density, elongated, magnetically-aligned structures (Goldsmith et al. 2008; Miville-Desch{\^e}nes et al. 2010; Palmeirim et al. 2013; Alves de Oliveira et al. 2014; Cox et al. 2016; Malinen et al. 2016; Panopoulou et al. 2016). Through a series of numerical experiments, Tritsis \& Tassis (2016) have demonstrated that the most probable formation mechanism for striations is that of compressible fast magnetosonic waves. In this physical picture, magnetic pressure waves compress the gas creating over-densities that align parallel to the magnetic field. A prediction from this physical model has been recently confirmed through the discovery of normal modes in the striations of the isolated cloud Musca (Tritsis \& Tassis 2018). Normal modes in Musca are established due to the trapping of magnetosonic waves by sharp gradients in their propagation speed between the cloud and the ambient medium. 

Magnetic pressure is dominant over thermal and turbulent pressures in regions with fibers (Heiles 1989; Dickey \& Lockman 1990). The dimensionless properties (i.e. the plasma $\beta$, defined as the ratio of thermal pressure to magnetic pressure, and the \Alf~Mach number) between striations and fibers are approximately the same. Thus, striations and fibers are likely to share the same formation mechanism. 

\begin{figure*}
\includegraphics[width=2.0\columnwidth, clip]{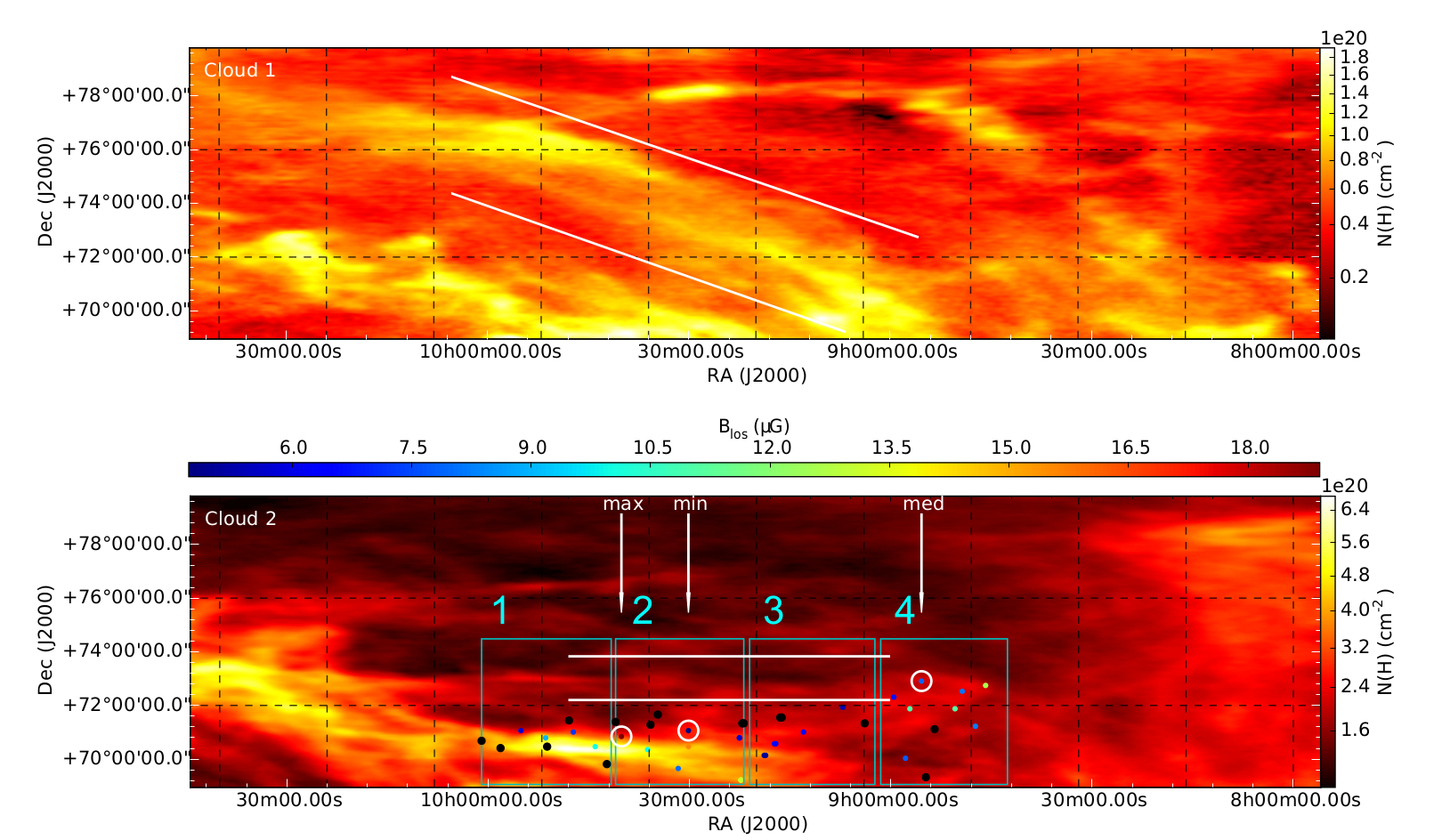}
\caption{Upper panel: column density of far cloud (Cloud 1). Lower panel: column density of near cloud (Cloud 2). On the column density map of Cloud 2 we overplot with colour-coded dots the LOS component of the magnetic field from Zeeman measurements (Myers et al. 1995). The values of the LOS magnetic field from these measurements are given in the horizontal colorbar. Black points are non-detection Zeeman measurements. All Zeeman detections yield positive values. White lines in both panels show the approximate direction of fibers. White arrows mark the positions where the value of the LOS magnetic field is minimum, maximum and equal to the median value from all measurements. The cyan rectangles annotated with numbers mark 4 regions where we performed our analysis individually (see text).  
\label{Combo}}
\end{figure*}

\begin{figure}
\includegraphics[width=1.0\columnwidth, clip]{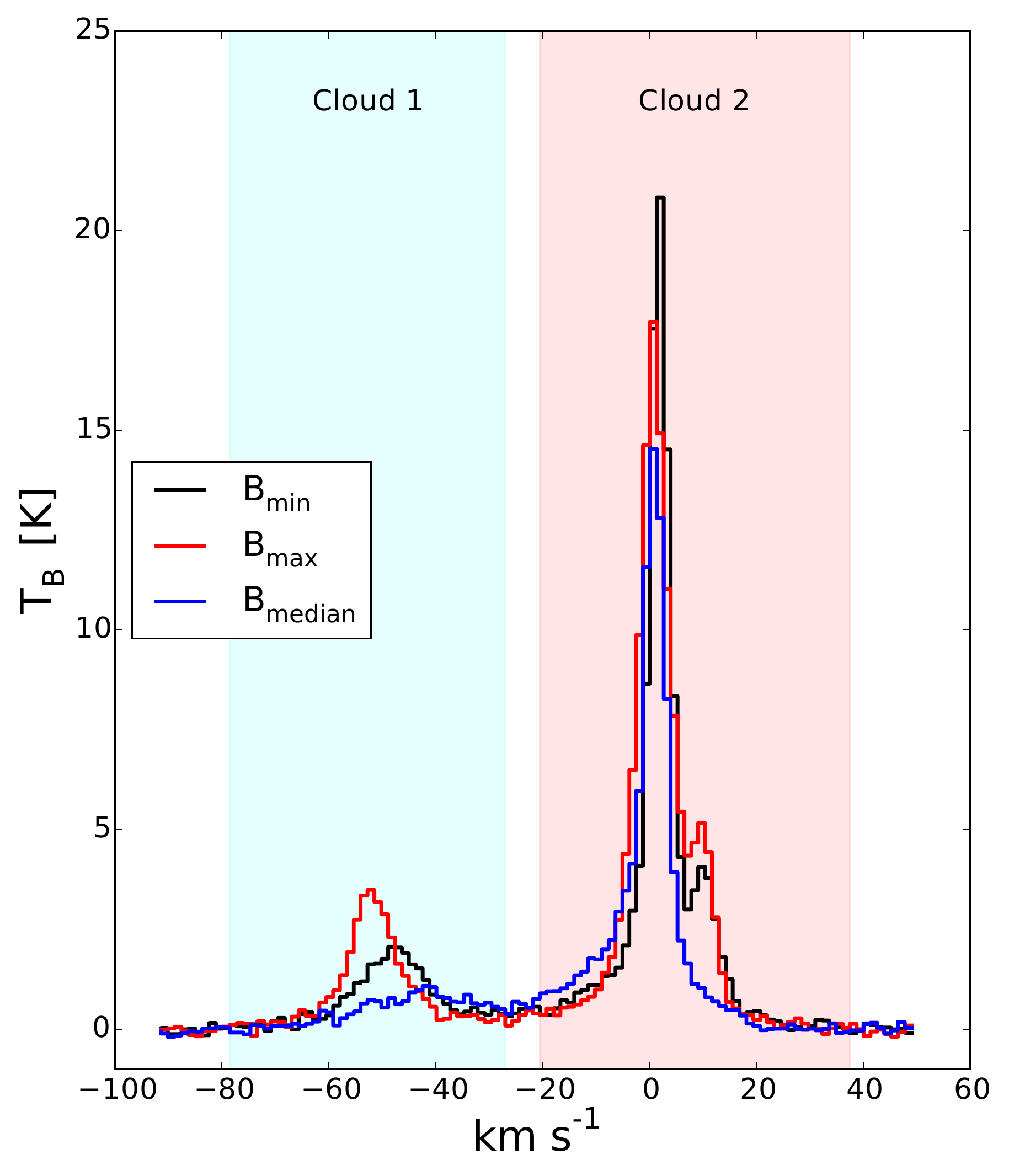}
\caption{Spectral lines from the sight-lines where the Zeeman magnetic field measurement is maximum, minimum and equal to the median value of all measurements (marked with white circles in the bottom panel of Figure~\ref{Combo}). The cyan and pink regions mark the velocity range we considered for producing the column density of Cloud 1 and Cloud 2, respectively.
\label{Slines}}
\end{figure}

Tritsis et al. (2018) recently developed a method for estimating the strength of the POS magnetic field through a combined analysis of column density and velocity centroid maps of striations. The method can be applied to individual clouds in different velocity slices and can thus be used to tomographically map the POS magnetic field. Zeeman measurements of the line-of-sight (LOS) magnetic field can also be taken in different velocity channels (e.g. Heiles \& Troland 2004). Apart from the strength of the LOS magnetic field, Zeeman measurements also yield its direction with the astronomical convention being that positive magnetic fields point away from the observer provided that the Stokes V is computed as the excess of right-circular polarization over the left-circular polarization (for a review on Zeeman observations conventions we refer the reader to Robishaw 2008). Combined, the two measurements can yield the total strength of the magnetic field, its angle with respect to the LOS and its direction in different clouds along a sight-line. 

Complete knowledge of the 3D structure of the magnetic field of the Galaxy would be extremely important for studies of the interstellar medium (in the context of interstellar cloud formation and evolution, star formation, and galaxy formation and evolution); but also for cosmic-ray propagation studies (e.g. Orlando \& Strong 2013; Planck Collaboration et al. 2016b; Pierre Auger Collaboration et al. 2017a; Magkos \& Pavlidou 2018). At the highest energies ($\geqslant$ $10^{19}$ eV), where cosmic rays despite being deflected by the magnetic field are not completely isotropized, knowledge of the 3D structure of the magnetic field could be used to reconstruct the path of individual cosmic rays through the Galaxy. Thus, by recovering the original velocity direction of cosmic rays before entering the Galaxy, such studies could elucidate the origin of observed cosmic-ray hotspots (Abbasi et al. 2014; Pierre Auger Collaboration et al. 2017b; Abbasi et al. 2018) and provide electromagnetic constraints on cosmic-ray composition, independent of particle physics (Pavlidou \& Tomaras 2018). Such a cosmic-ray hotspot is observed towards Ursa Major, making this region ideal for tomographically mapping the magnetic field. At the same time, knowledge of the 3D structure of the magnetic field and dust alignment and emission properties (for a recent review on grain alignment see Andersson et al. 2015) could be used to predict the expected polarization in regions away from the Galactic plane. Such information would be invaluable for cosmic microwave background polarization (CMB) experiments and the complications that arise due to dust foregrounds (BICEP2/Keck Collaboration et al. 2015; Planck Collaboration et al. 2016c; Tassis \& Pavlidou 2015).

In this paper, we provide the first estimate of the strength of the POS magnetic field in two clouds along the same sight-line towards the boundaries of Ursa Major with Camelopardalis and Draco. For the cloud located closer to Earth, we combine our results with Zeeman measurements from the literature (Goodman et al. 1994; Myers et al. 1995). To measure the POS component of the magnetic field we apply the method by Tritsis et al. (2018), developed for molecular cloud striations. In \S\ref{method} we summarize the theoretical background described in detail in Tritsis et al. (2018). The observations we used for our analysis are presented in \S\ref{observations}. In \S\ref{analysis} we describe our analysis and in \S\ref{Bfield} we employ these observations to estimate the magnetic field. We discuss our results in \S\ref{discuss} and summarize in \S\ref{sum}.
%=-=-=-=-=-=-=-=-=-=-=-=-=-=-=-=-=-=-=-=-=-=-=-=-=-=-=-=-=-=-=-=-=-=-=%
%=-=-=-=-=-=-=-=-=-=-=-=-=-=-=-=-=-=-=-=-=-=-=-=-=-=-=-=-=-=-=-=-=-=-=%

%=-=-=-=-=-=-=-=-=-=-=-=-=-=-=-=-=-=-=-=-=-=-=-=-=-=-=-=-=-=-=-=-=-=-=%
%=-=-=-=-=-=-=-=-=-=-=-=-=-=-=-=-=-=-=-=-=-=-=-=-=-=-=-=-=-=-=-=-=-=-=%

\section{Method}\label{method}
By linearising the continuity equation, the induction equation of ideal magnetohydrodynamics (MHD), and assuming that the displacement of the gas is described by a plane-wave solution (for details see Tritsis et al. 2018) we obtain:
\begin{equation}\label{vel}
\delta v = -\sum_n i A_n\omega_n e^{i(\mathbf{k_nr}-\omega_n t)}
\end{equation}
\begin{equation}\label{cd}
\delta N_{H} = N^0_{H}\sum_n A_n k_n ie^{i(\mathbf{k_nr}-\omega_n t)}
\end{equation}
\begin{equation}\label{induct}
\delta B = B_0\sum_n A_n k_n ie^{i(\mathbf{k_nr}-\omega_n t)}\
\end{equation}
where $\omega$ is the frequency of the waves, $A$ is the amplitude, $k$ is the wavenumber and the subscript $n$ is used to denote different waves. Equation~\ref{vel} is written in the frame of reference of the cloud such that the perturbation in velocity ($\delta v$) in the perpendicular-to-the-magnetic-field direction is measured with respect to the bulk velocity ($v_0$) of the cloud. In Equation~\ref{cd}, $N\rm{^0_{H}}$ is the mean column density and $\delta N\rm{_{H}}$ is the perturbation in column density. Similarly, in Equation~\ref{induct}, $B_0$ and $\delta B$ are the mean magnetic field and the perturbation in the magnetic field, such that $B_{tot} = B_0 + \delta B$ and $N\rm{^{tot}_{H}} = $  $N\rm{^0_{H}} +$ $\delta N\rm{_{H}}$.

In Equations~\ref{vel},~\ref{cd} \&~\ref{induct}, the exponential part is the same and only the coefficients change. Thus, the spatial power spectra of the observed velocity, column density and magnetic field should peak at the same wavenumbers. Observationally, such power spectra can be computed by considering cuts perpendicular to the long axis of fibers. The power in the column density power spectra $\vert P_{N^1_{H}} \vert^2$ would be $\vert N^0_{H} A_n k_n\vert^2$ and the power in velocity centroid power spectra $\vert P_{v_1} \vert^2$ would be $\vert\omega_n A_n\vert^2$. We introduce the parameter $\Gamma$, defined in Tritsis et al. (2018) as the square root of the ratio of the power of velocity power spectra over the power of column density power spectra:
\begin{equation}\label{gamma}
\Gamma_n = \sqrt{\frac{\vert P_{v_1} \vert^2}{\vert P_{N^1_{H}} \vert^2}} = \frac{\omega_n}{k_n}\frac{1}{N^0_{H}}
\end{equation}
The term $\omega_n/k_n$ in the latter Equation can be substituted from the dispersion relation of fast magnetosonic waves (for a review on MHD waves see Spruit 2013):
\begin{equation}\label{dispers}
\frac{\omega_n}{k_n} = \sqrt{\frac{1}{2} \Big[(v_A^2 + c_s^2) + \sqrt{(v_A^2 + c_s^2)^2 - 4v_A^2c_s^2 \rm{cos \phi} } ~\Big] }
\end{equation}
where $v\rm{_A}$ is the \Alf~speed (defined as $v\rm{_A} = B_0/\sqrt{4\pi\rho_0}$, where $\rho_0$ is the mean density), $c\rm{_s}$ is the sound speed and $\phi$ is the angle between the wavevector and the magnetic field such that, when $\phi = \pi /2$, the waves propagating in the medium are magnetic pressure waves exactly perpendicular to the magnetic field. In Equation~\ref{dispers} the sound speed $c\rm{_s}$ can be ignored since $v\rm{_A}^2 \gg c\rm{_s}^2$. Additionally, by considering only the waves that propagate perpendicular to the magnetic field, we obtain for the dispersion relation of fast magnetosonic waves that $\omega_n/k_n \approx v\rm{_A}$. Finally, by combining the latter equation with Equation~\ref{gamma}, we obtain:
\begin{equation}\label{mainEqIn}
\Gamma_n = \frac{v_A}{N^0_{H}}  = \frac{B_0}{\sqrt{4\pi\rho_0}}\frac{1}{N^0_{H}} \Rightarrow B_0 = \Gamma_n N^0_{H} \sqrt{4\pi \rho_0}.
\end{equation}
Waves propagating in directions other than perpendicular to the field might as well be present and this will create a non-zero spread in the distribution of orientation angles of fibers. However, by considering cuts perpendicular to fibers, only the waves that propagate perpendicular to the magnetic field are probed in our analysis. For a full description of the simplifications made and a thorough justification of the assumptions entering this derivation, we refer the reader to Tritsis et al. (2018). Equation~\ref{mainEqIn} is valid when the magnetic field lies entirely on the POS and the LOS component is zero. When the magnetic field lies at an angle $\theta$ with respect to the POS, the \textit{intrinsic} perturbations ($\delta v$) in the velocity component perpendicular to the magnetic field will be \textit{observed} as $\delta v\rm{cos}\theta$ and the power in the spectra of velocity centroids perpendicular to fibers will be $\vert\omega_n A_n \rm{cos}\theta \vert^2$. Thus, by taking projection effects into account, we find:
\begin{equation}\label{mainEq}
B_0\mathrm{cos}\theta = B^{pos}_0 = \Gamma_n N^0_{H} \sqrt{4\pi \rho_0}.
\end{equation} 
Consequently, the method can only probe the POS component of the magnetic field and the angle $\theta$ can be found in combination with measurements of the LOS component via Zeeman observations.

%=-=-=-=-=-=-=-=-=-=-=-=-=-=-=-=-=-=-=-=-=-=-=-=-=-=-=-=-=-=-=-=-=-=-=%

\section{Data}\label{observations}

For our analysis we use publicly available data from the EBHIS survey (Winkel et al. 2016) and published Zeeman measurements (Goodman et al. 1994; Myers et al. 1995). The spectral resolution of the EBHIS data is 1.29 km $\rm{s^{-1}}$, the angular resolution is $\sim$ 11 arcmins, the main beam sensitivity is 1.434 $\rm{K~Jy^{-1}}$ and the data are stray-radiation corrected. The analysis of the Zeeman observations by Myers et al. (1995) is consistent with all astronomical conventions. The two clouds are located at $\sim$200 $\rm{pc}$ and $\sim$1 $\rm{kpc}$ respectively (see Appendix~\ref{distance}).

\section{Analysis}\label{analysis}

Along the same sight-line as the Zeeman observations there are two clouds (see Figure~\ref{Combo}). We obtained the column density of each cloud assuming optically thin conditions:
\begin{equation}\label{cdDeriv}
N\rm{_{H}} = 1.823\times 10^{18} \,\rm{cm^{-2}} \times \int \mathit{T_B} ~[\rm{K}]~\, \mathit{dv} ~[\rm{km \, s^{-1}}]
\end{equation}
(Dickey \& Lockman 1990; Chengalur et al. 2013) where $\int T\rm{_B} dv$ is the velocity-integrated brightness temperature. Our results are shown in Figure~\ref{Combo}. The upper and lower panels show the column density maps of the cloud located further away from Earth (henceforth ``Cloud 1") and the cloud located closer to Earth (henceforth ``Cloud 2"), respectively. The colour-coded dots overplotted on the lower panel are the results from the Zeeman observations by Myers et al. (1995). Non-detections are overplotted as black dots. In order to investigate whether the magnetic field changes direction we separate Cloud 2 in 4 regions shown with the green rectangles in the bottom left panel in Figure~\ref{Combo}. We then apply the method by Tritsis et al. (2018) in each of these regions individually and compare our results with the Zeeman measurements from each region. In Figure~\ref{Slines} we plot 3 spectral lines for 3 different lines-of-sight: where the magnetic field from the Zeeman measurements is minimum, where it is maximum and where it equals the median value of all measurements (marked with white circles and arrows in the lower panel of Figure~\ref{Combo}). The light blue shaded region marks the velocity range of Cloud 1 and the pink shaded region marks the velocity range of Cloud 2. 

In order to apply the method by Tritsis et al. (2018) we created the first-moment map of velocity of each cloud by fitting one or two Gaussian distributions to each of the spectral lines of each pixel and computing the velocity of each cloud as a weighted average of the two Gaussians (these maps are presented in Appendix~\ref{fmm}). Gaussian fits were only performed to features in the spectra with signal more than 5 times the noise level. By considering the mean and standard deviation of the first-moment map of each cloud, we find that Cloud 1 is centred at -43 $\pm$ 6 km $\rm{s^{-1}}$ and Cloud 2  at -0.5 $\pm$ 3.7 km $\rm{s^{-1}}$.  

We consider cuts perpendicular to fibers in both the column density map and the first-moment map of velocity and compute their power spectra. From the square root of the ratio of powers of velocity and column density spectra at the same peaks we compute the parameter $\Gamma$. In order to ensure that a peak in the velocity power spectrum corresponds to the same spatial scale as a peak in the column density power spectrum (as opposed to another spatial scale from an adjacent peak), we compare their respective wavenumbers. We require that the difference in the values of wavenumbers of two spectra is not larger than 15\% the value of the wavenumber in column density in the location of the peak. We further apply the criterion that the parameter $\Gamma$ is only computed from peaks that have powers greater than 10\% the maximum power in the power spectrum of each cut\footnote{We made sure that our results are not affected by the choice of these two numbers (i.e. 15\% and 10\%) by varying them by a factor of 2, above and bellow their fiducial values.}. Additionally, we require that a power spectrum must have in total more than 2 peaks with power greater that 10\% the maximum power. In this manner, we avoid spurious peaks and ensure that failed Gaussian fits from the process of producing the first-moment map do not affect our results. Finally, we estimate the errors in measuring the wavenumbers and the powers in each power spectrum (see Appendix~\ref{errors}) which we then use in our analysis.

\begin{figure}
\includegraphics[width=1.0\columnwidth, clip]{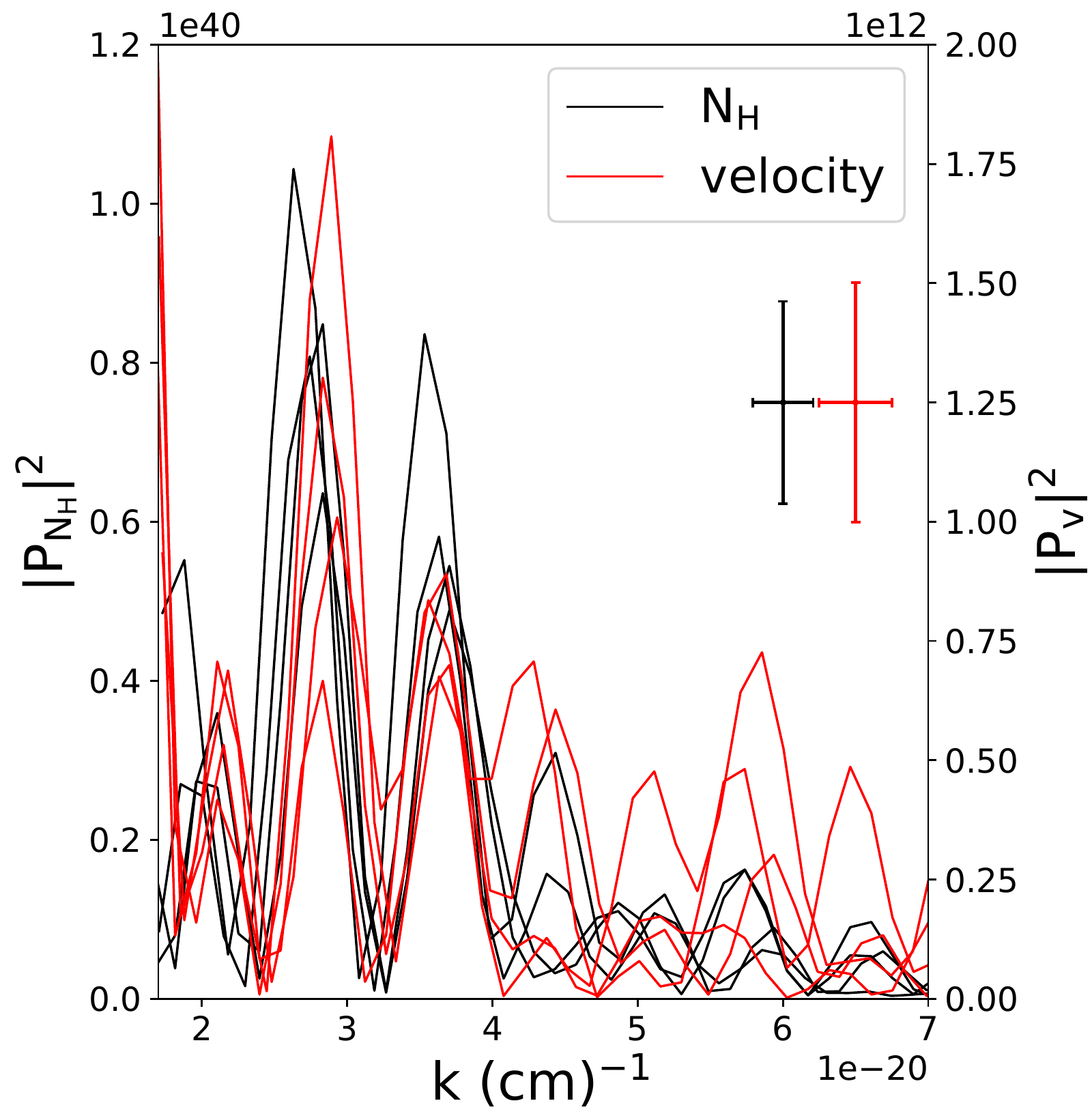}
\caption{Example of power spectra for Cloud 1 of column density cuts perpendicular to fibers (black lines) and velocity centroid cuts perpendicular to fibers (red lines). The black and red errorbars show the typical 1$\sigma$ errors in determining the power and the wavenumber of each column density and velocity power spectrum (see Appendix~\ref{errors}). The power spectra peak at approximately the same wavenumbers, in agreement with the theoretical predictions of \S~\ref{method}. %The inset figure shows the parameter $\Gamma$ as a function of the wavenumber for each of the power spectra. Different symbols denote our results from each individual set (velocity and column density) of spectra resulting from different cuts perpendicular to fibers.
\label{PSobs1}}
\end{figure}

\begin{figure}
\includegraphics[width=1.0\columnwidth, clip]{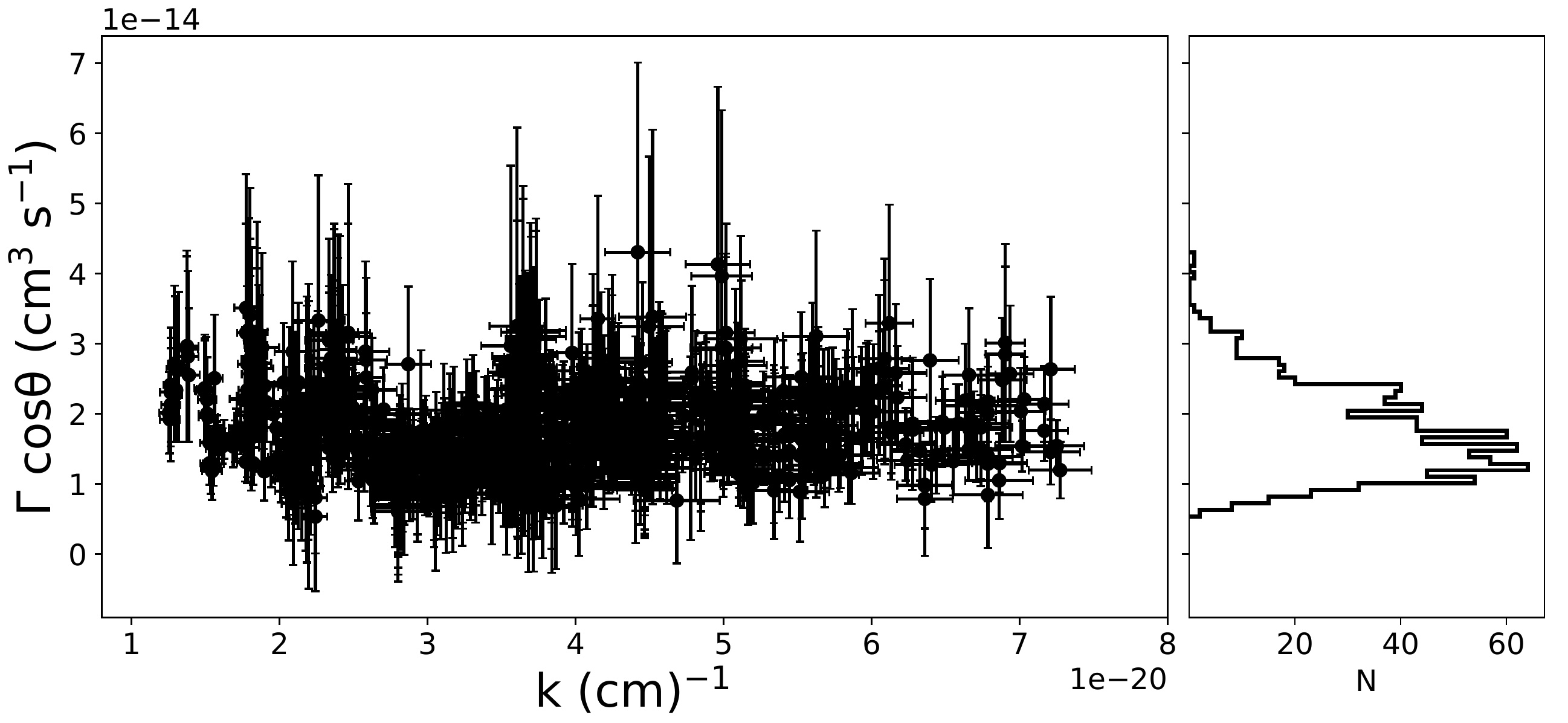}
\caption{Left panel: the parameter $\Gamma$ as a function of the wavenumber from the power spectra from all cuts perpendicular to fibers. The errorbars show the 1$\sigma$ uncertainty, computed from error propagation from the errors in determining the power and the wavenumber of each column density and velocity power spectrum (Figure~\ref{PSobs1}). In agreement with the theoretical predictions of \S~\ref{method}, despite the scatter, there is no correlation of the parameter $\Gamma$ with the wavenumber. Right panel: Distribution of the values of the parameter $\Gamma$.
\label{GkB}}
\end{figure}

\begin{figure}
\includegraphics[width=1.0\columnwidth, clip]{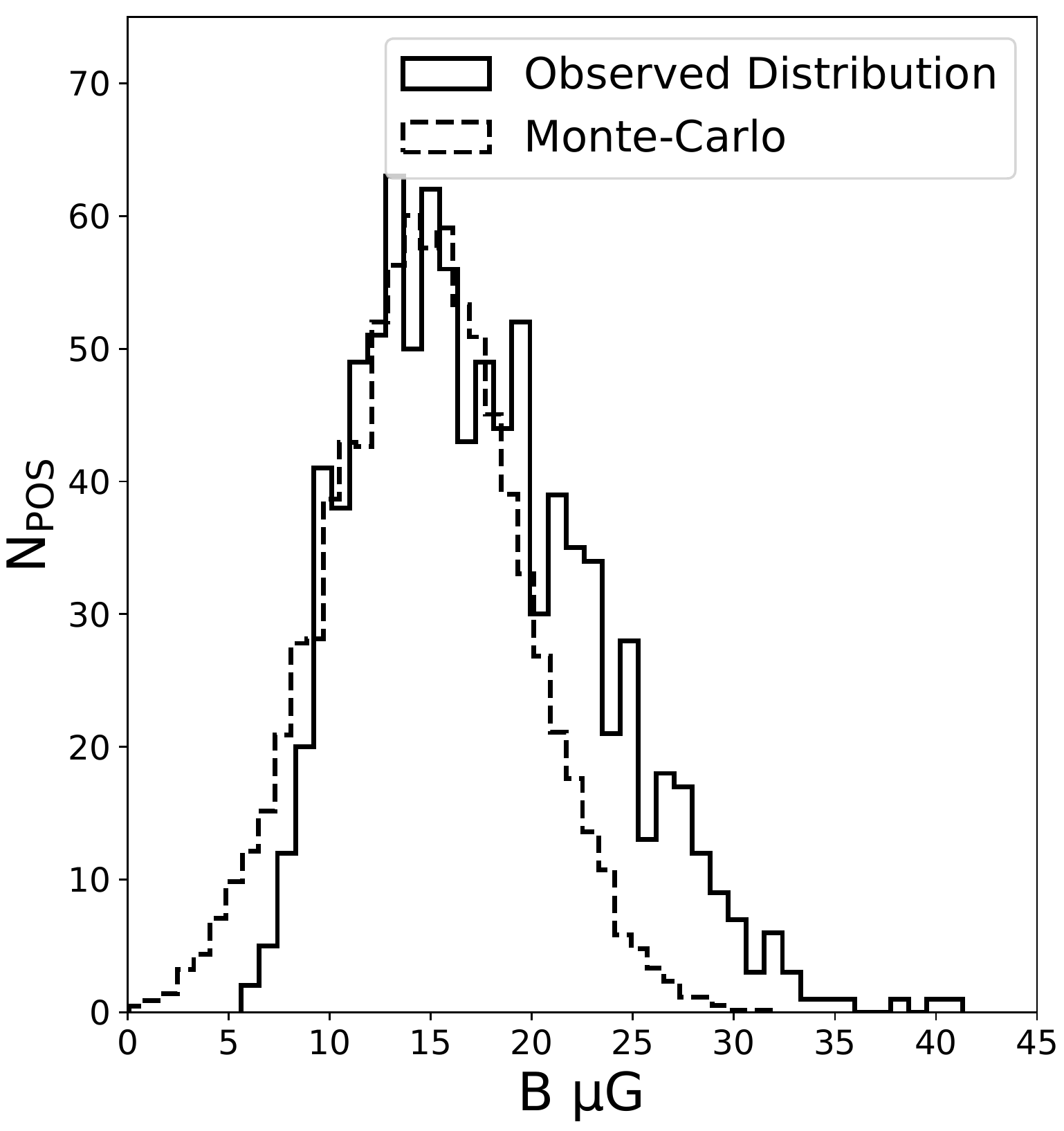}
\caption{Distribution of POS magnetic field values (solid black line), derived from the values of the parameter $\Gamma$ of the points shown in Figure~\ref{GkB}. The dashed black line shows a Monte Carlo estimate of the normalised distribution of POS magnetic field values, computed from the mode of the distribution of the parameter $\Gamma$ (see left panel of Figure~\ref{GkB}) and typical errors. 
\label{Bdistro1}}
\end{figure}

\section{Results}\label{Bfield}

\subsection{Cloud 1}

Examples of column density and velocity power spectra are shown in Figure~\ref{PSobs1} (black and red lines respectively). The black and red crosses show typical errors in the column density and velocity power spectra respectively (see Appendix~\ref{errors}). The power spectra of velocity and column density have peaks at approximately the same spatial frequencies. The variation of power between different peaks is similar in both the velocity and column density power spectra. For example, in both the velocity and column density power spectra, the second peak has the maximum power, the third peak has the second most power and the first peak has the least power amongst these three peaks. Thus, the results presented in Figure~\ref{PSobs1} are in agreement with the theoretical predictions from \S~\ref{method}. 

%For example, from a set of velocity and column density spectra shown in Figure~\ref{PSobs1}, the parameter $\Gamma$ at the first peak ($k \sim 2 \cdot 10^{-20}~ \rm{cm^{-1}}$ ) will be $\sim ~1.2\cdot10^{-14}$ and at the second peak ($k \sim 2.8 \cdot 10^{-20}~ \rm{cm^{-1}}$ ) it will be $\sim ~1.3\cdot10^{-14}$.  In the inset figure we plot the parameter $\Gamma$ as a function of the wavenumber for each set of peaks that fulfil all our criteria from the power spectra shown.  

\begin{figure}
\includegraphics[width=1.0\columnwidth, clip]{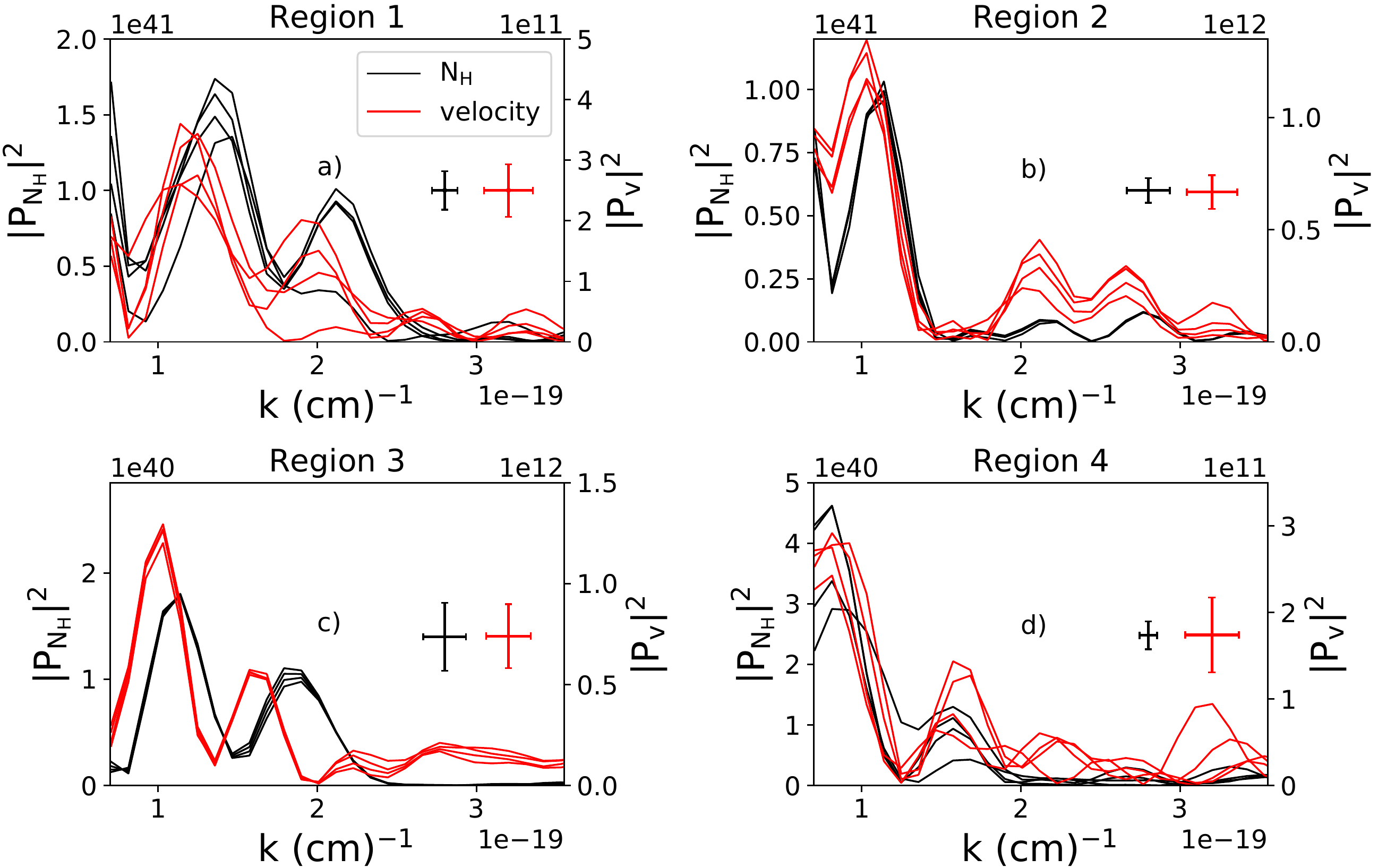}
\caption{Power spectra of velocity centroid and column density cuts (red and black lines respectively) in the four regions of Cloud 2. Errorbars are the same as in Figure~\ref{PSobs1}. As in Cloud 1, the power spectra peak at approximately the same wavenumbers. 
\label{PS2}}
\end{figure}

In the left panel of Figure~\ref{GkB} we show the parameter $\Gamma$ as a function of the wavenumber in each peak. Errorbars are computed from the errors in the column density and velocity power spectra (see Figure~\ref{PSobs1}) through error propagation. Since the variation of power between different wave-modes follows the same pattern for velocity and column density power spectra (see Figure~\ref{PSobs1}), the ratio of their powers remains approximately constant, in agreement with the theoretical predictions of \S~\ref{method}. In the right panel of Figure~\ref{GkB} we show the distribution of the values of the parameter $\Gamma$. The mode of this distribution along with the 16th and 84th percentiles is $1.4^{+10}_{-3}\times10^{-14}~\rm{cm^3 ~s^{-1}}$. In Figure~\ref{Bdistro1} we plot the distribution of magnetic field values (solid black line) derived from the values of the parameter $\Gamma$, the density of the cloud and the mean column density of each respective column density cut. We assume that the number density of the cloud is 10 $\rm{cm^{-3}}$. The mode with the 16th and 84th percentiles is $15^{+8}_{-3}~\rm{\upmu G}$. Thus, the derived magnetic field value for Cloud 1 is larger than typical values measured in H\textsc{I} clouds and the values derived from models (Heiles 1989; Haverkorn 2015). We further investigate whether the spread of the distribution shown with the solid black line in Figure~\ref{Bdistro1} originates from intrinsic variations of the magnetic field or from uncertainties of the method by performing a simple Monte Carlo simulation. We draw $10^6$ values for the parameter $\Gamma$ from a Gaussian distribution where the mean of the Gaussian is the mode of the observed distribution of the parameter $\Gamma$ (see right panel of Figure~\ref{GkB}) and as $\sigma$ we adopt a typical value of the error (see left panel of Figure~\ref{GkB}). We plot our results with the dashed line in Figure~\ref{Bdistro1}. The two distributions are in good agreement and any differences can be attributed to the fact that neither the parameter $\Gamma$ nor its errors follow a Gaussian distribution. However, what this simple Monte Carlo simulation shows is that the POS value of the magnetic field in Cloud 1 does not vary much and that any intrinsic variations are hidden under the uncertainties of the method.

%The constant ratio of powers as a function of the wavenumber is not an effect of stacking the results from many peaks. This is demonstrated in the inset figure of Figure~\ref{PSobs1} where we show the parameter $\Gamma$ as a function of the wavenumber from each individual set of spectra. 

\begin{figure}
\includegraphics[width=1.0\columnwidth, clip]{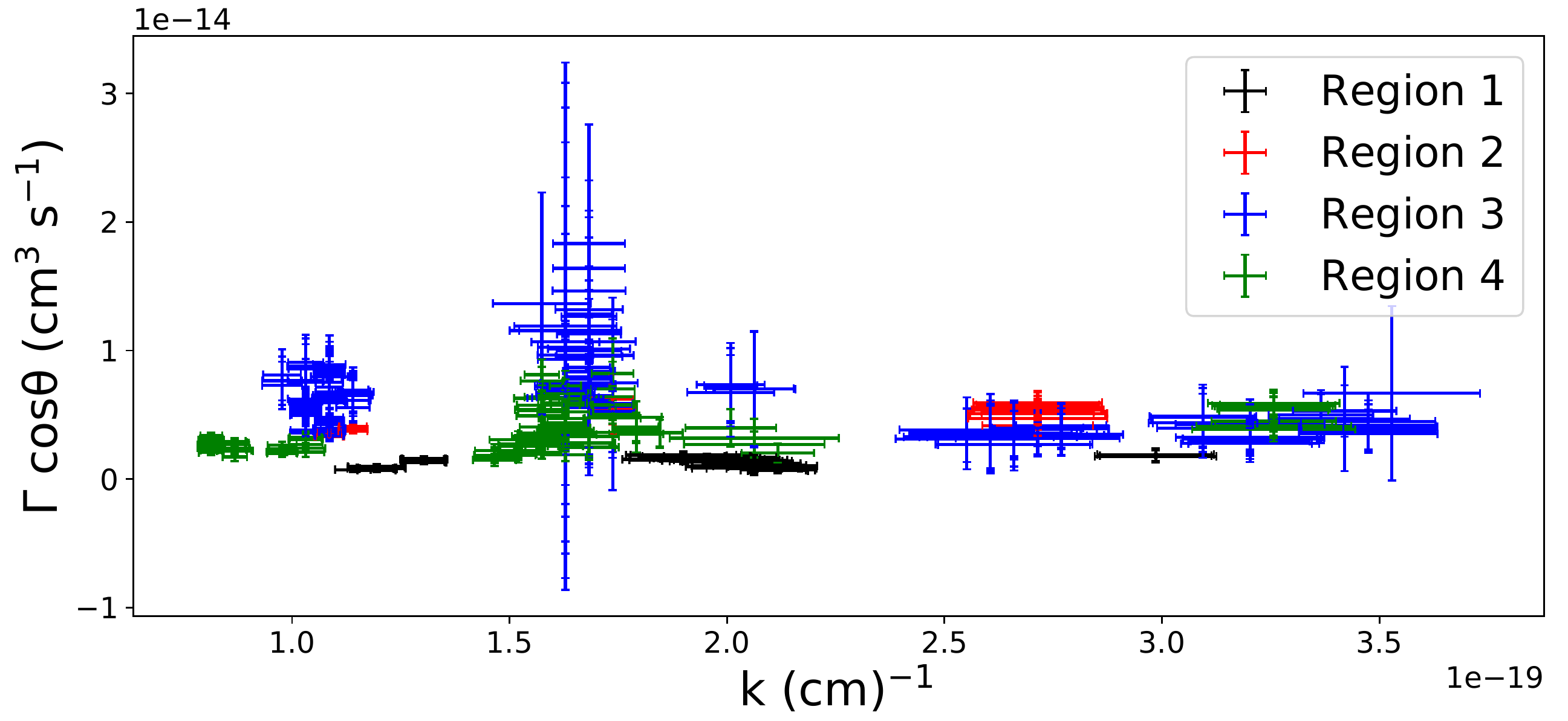}
\caption{The parameter $\Gamma$ as a function of the wavenumber for each of the four regions of Cloud 2. Errorbars are the same as in Figure~\ref{GkB}.
\label{Gk}}
\end{figure}

\subsection{Cloud 2}

We repeat our analysis for Cloud 2 which is located closer to Earth and for which additional Zeeman measurements are available. Since for this cloud we have information about the LOS component of the magnetic field and in order to investigate potential changes in the direction of the magnetic field we separate Cloud 2 into 4 regions (from 1 to 4 as shown in Figure~\ref{Combo}). In Figure~\ref{PS2} we show examples of power spectra of velocity and column density cuts from all regions. In agreement with the theoretical analysis of \S~\ref{method} and the results for Cloud 1 the power spectra peak at approximately the same wavenumbers and the variations in the power of velocity and column density follow each other well.

\begin{figure}
\includegraphics[width=1.0\columnwidth, clip]{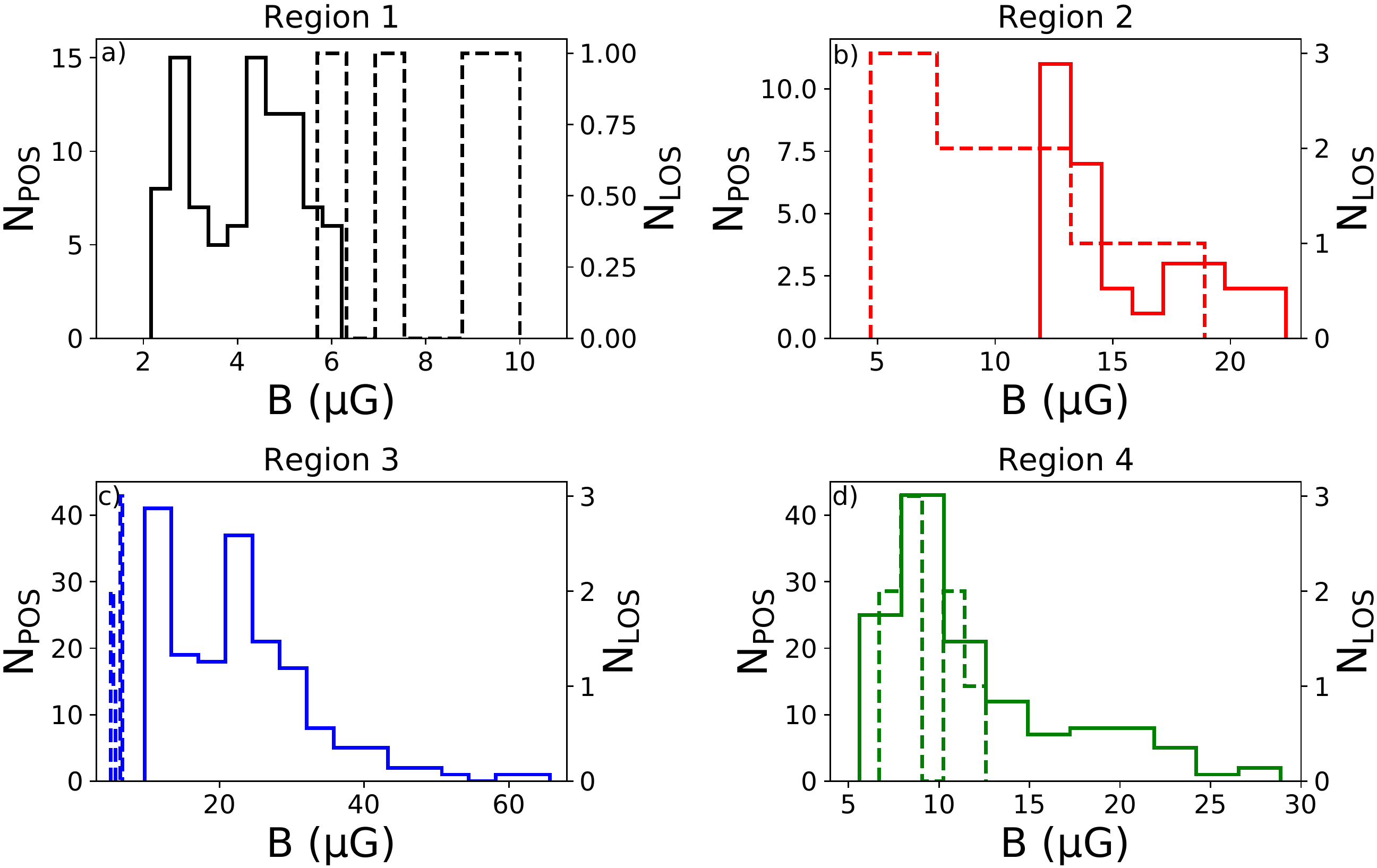}
\caption{Magnetic field distributions for each of the four regions of Cloud 2. The solid distributions correspond to the left y-axis ($\rm{N_{pos}}$) and show our results for the POS component of the magnetic field. Dashed lines show the distributions of the LOS magnetic field values for each of the four regions measured via Zeeman observations and correspond to the right y-axis ($\rm{N_{los}}$).  Especially for region 3 (panel c), the derived magnetic field values are  unusually high. 
\label{Bcl2}}
\end{figure}

In Figure~\ref{Gk} we plot the parameter $\Gamma$ as a function of the wavenumber for each of the four regions of Cloud 2. Results for region 1 are shown with black squares, for region 2 with red dots, for region 3 with blue stars and for region 4 with green crosses. For each of the four regions the value of the parameters $\Gamma$ remains approximately constant for different wavenumbers. This is again in agreement with the theoretical considerations of \S~\ref{method} and the results for Cloud 1. 

In Figure~\ref{Bcl2} we plot the distributions of magnetic field values derived from the points shown in Figure~\ref{Gk} for region 1 (solid black line), region 2 (solid red line), region 3 (solid blue line) and region 4 (solid green line) for Cloud 2. Dashed lines show the LOS components of the magnetic field for the same regions from Zeeman observations by Myers et al. (1995). As in Cloud 1, the POS magnetic field values are computed from Equation~\ref{mainEq} where for $N^0_{H}$ we use the mean column density of each respective column density cut. From the adopted distance of Cloud 2 (see Appendix~\ref{distance}), the mean column density, the size of the region and assuming that the LOS dimension is equal to the size of the region projected on the POS, the number density of Cloud 2 is $\sim$ 10 $\rm{cm^{-3}}$. The mode with the 16th and 84th percentiles is $4^{+1}_{-1}~\rm{\upmu G}$ for region 1, $14^{+7}_{-2}~\rm{\upmu G}$ for region 2, $17^{+14}_{-4	}~\rm{\upmu G}$ for region 3 and $10^{+8}_{-2}~\rm{\upmu G}$ for region 4. In reality, the density of Cloud 2 may be much higher (see \S~\ref{discuss}). Thus, the values quoted above should be interpreted as lower limits. 

In Figure~\ref{Combo} there is evidence of a column density gradient moving from region 1 to region 4 and thus region 1 might be denser than region 4. Assuming that the density in region 1 is a factor of two larger than that of region 3 or 4, the mode of the POS magnetic field strength will be $5^{+2}_{-1}~\rm{\upmu G}$. Thus, a density gradient from region 1 to region 4 cannot explain the differences seen in the distributions of the magnetic field between different regions. A more plausible explanation is that the magnetic field changes direction between different regions.

\begin{figure*}
\includegraphics[width=2.0\columnwidth, clip]{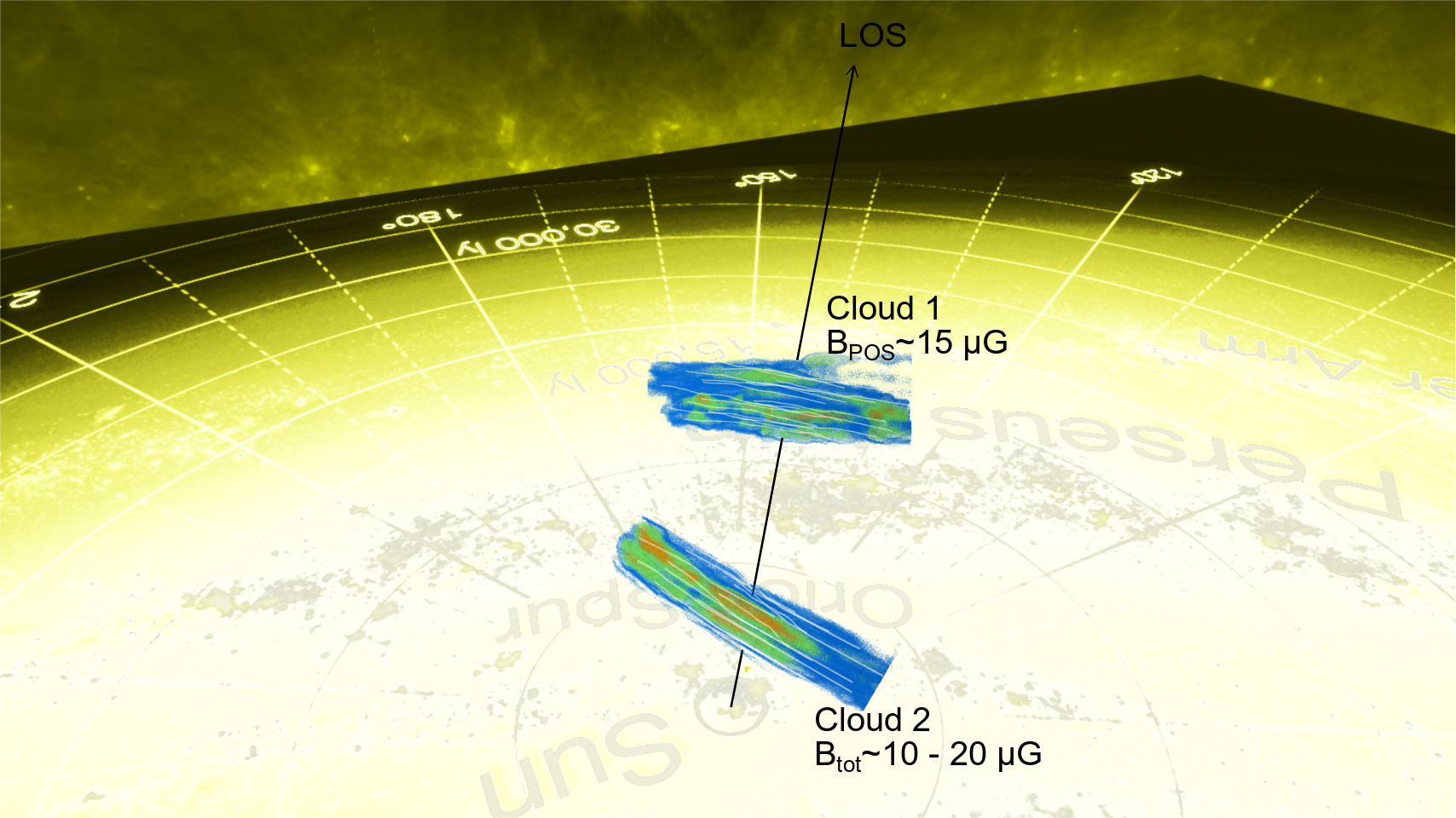}
\caption{3D view of the two clouds and their location in the Galaxy. The magnetic field of each cloud is shown with the white lines. The black arrow shows the LOS where the two clouds are observed (produced with the Space Nebula Plugin for Unreal Engine 4).  
\label{Galaxy3D}}
\end{figure*}

In Figure~\ref{Galaxy3D} we summarize our results with a 3D view of the clouds inside the Galaxy. In Table~\ref{table} we summarize the values for the POS component of the magnetic field 

\begin{table}[]
\centering
\caption{Estimated values of the magnetic field in the two clouds. For Cloud 2 we summarize the magnetic field value for each of the four regions where we performed our analysis. The values for the LOS component of the magnetic field given are the mean and standard deviation of the dashed distributions shown in Figure~\ref{Bcl2}.}
\begin{tabular}{llll}
\hline\hline
& & $\rm{B_{pos}}~(\rm{\upmu G})$ & $\rm{B_{los}}~(\rm{\upmu G})$ \\
\hline
Cloud 1 & & $15^{+8}_{-3}$ &  Zeeman data N/A \\
\hline
Cloud 2 & region 1 & $4^{+1}_{-1}$ & $8\pm 2$ \\
& region 2 & $14^{+7}_{-2}$ & $10\pm 5$ \\
& region 3 & $17^{+14}_{-4	}$ & $6\pm 1$ \\
& region 4 & $10^{+8}_{-2}$ & $9\pm 2$  \\
\hline\hline
\end{tabular}\label{table}
\end{table}

\section{Discussion}\label{discuss}

In Cloud 2 the distributions of the values of the POS and LOS components of the magnetic field are in fair agreement. On the other hand, significant variation is observed between regions. In region 1, the LOS component is higher than the POS component ($\sim$ 24$^\circ$ with respect to the LOS) whereas in region 2 the opposite trend is observed with the POS component being higher than the LOS component ($\sim$ 54$^\circ$ with respect to the LOS) and the two distributions overlap. In region 3 the POS component becomes much higher than the LOS component ($\sim$ 71$^\circ$ with respect to the LOS) and with no overlap between the two distributions and finally, in region 4 the two components are approximately the same and equal to $\sim$ 10 $\rm{\upmu G}$ ($\sim$ 51$^\circ$ with respect to the LOS). Such a behaviour can be potentially explained with the mean magnetic field following an \Alf~wave primarily polarized in the LOS direction. However, more Zeeman measurements are required in order to verify this hypothesis, or identify other potential interpretations. 

According to the Monte-Carlo simulations by Chengalur et al. (2013), departures from isothermality do not significantly affect the derived column density which is computed within a factor of 2 of the true value. However, in order to achieve that accuracy, the optical depth has to be known from absorption studies. Even if such studies were available for the two clouds analysed in this paper, knowledge of the LOS dimension is still required in order to better constrain their density (that enters the calculations) from the column density. The mean density of the cold neutral medium is $\sim$ 60 $\rm{cm^{-3}}$ (Heiles \& Troland 2003). Based on the column densities derived for the clouds, a fraction of their mass might also be in the form of molecular hydrogen (Planck Collaboration et al. 2011). In fact, molecular gas has been observed at the same location and velocity of Cloud 2 (Magnani et al. 1996). Thus, the value of $10~\rm{cm^{-3}}$ adopted for the density of the two clouds should be considered as a lower limit. We emphasize here that this uncertainty in density enters the calculation of the POS component of the magnetic field in all methods (e.g. Davis 1951; Chandrasekhar \& Fermi 1953; Clark et al. 2014; Gonz{\'a}lez-Casanova \& Lazarian 2017). However, the POS component of the magnetic field scales as the square root of density. Thus, despite these uncertainties, the true value of the POS component of the magnetic field in each cloud would not be significantly affected by updated density estimates and would remain within a factor of two higher than the values quoted here.

\subsection{Physics of Fibers}

The upper panel of Figure~\ref{GkB} and Figure~\ref{Gk} represents the dispersion relation of fast magnetosonic waves derived from observations of fibers. Furthermore, the fact that velocity and column density power spectra peak at the same positions (Figures~\ref{PSobs1} \&~\ref{PS2}) supports our original hypothesis, that fibers, similarly to molecular cloud striations, are created from hydromagnetic waves.

Caldwell et al. (2017) analysed Planck polarization data and found that the parameter space required for MHD turbulence to account for observations is very limited. This is further supported by the fact that the ratio of the turbulent to ordered component of the magnetic field is found to be below unity (Planck Collaboration et al. 2016a; Planck Collaboration et al. 2016d; Planck Collaboration et al. 2018). In the case of Burgers turbulence (Burgers 1948), the power in the velocity power spectra scales as $P_v \propto k^{-2}$ (Solomon et al. 1987). Depending on the exact physical parameters and the type of turbulent forcing (solenoidal or compressible), the power in column density power spectra for supersonic clouds without self-gravity can scale from $P_{N_H} \propto k^{-1}$ for highly supersonic clouds up to $P_{N_H} \propto k^{-3.7}$ for slightly supersonic clouds (Federrath \& Klessen 2013). As a result, the ratio of powers of column density and velocity power spectra, or equivalently the parameter $\Gamma$, should depend on the wavenumber as $\Gamma \propto k^{a}$ with the spectral index $a$ in the range -1 to 1.7. From this range of values, only $a=0$ is consistent with the results presented in the upper panel of Figure~\ref{GkB} and in Figure~\ref{Gk}. Observationally, the velocity spectral index found for H\textsc{I} clouds at high galactic latitudes is $\sim$ 2 and that of column density is $\sim$ 1 (Chepurnov et al. 2010). From these observational results, the parameter $\Gamma$, should depend on the wavenumber as $\Gamma \propto k^{-1}$. However, it is unclear what the dependence of the parameter $\Gamma$ on the wavenumber would be in the case of turbulence, if the spectral indices were computed only in the direction perpendicular to field lines. Thus, although $\Gamma$ being independent of $k$ in the case of turbulence remains a possibility, the results presented in this paper can be naturally explained if fibers are created from fast magnetosonic waves.

Goodman et al. (1994) presented in graphical form the results from Zeeman measurements by Myers et al.(1995). They found that a profile of IRAS 100 $\upmu m$ flux and a profile of magnetic field measurements followed each other very well. This is in agreement with the theoretical predictions of \S~\ref{method} in which the column density, velocity centroids and magnetic field variations should all be correlated for a cut across fibers. This discussion along with our results for the values of the magnetic field in the two clouds should be taken into consideration for future numerical studies of the formation of molecular clouds and for the evolution of H\textsc{I} clouds (e.g. Inoue \& Inutsuka 2016; Gazol \& Villagran 2018).

\subsection{Implications for the nature and propagation of cosmic rays}

The values we derived for the magnetic field strength using our local, tomographic method are considerably larger than the values predicted by global Galactic magnetic field models that rely on fitting a mix of likely magnetic field components to line-of-sight integrated observables such as Faraday rotations, Synchrotron emission, or polarized dust emission. In particular, Galactic field models (Sun et al. 2008; Jansson \& Farrar 2012) estimate a magnetic field 3-5 times weaker than our estimate at the distance of the near cloud (200 pc), and a factor at least 5-8 times weaker than our estimate at the distance of the far cloud (1 kpc). The disagreement is even larger if we instead adopt the (higher) kinematic distances for the clouds (see Appendix~\ref{distance}), or if we adopt a higher value for the volume density. 

This result has particularly important implications since the sight-line we have examined is in the general direction of the northern ultra-high-energy cosmic-ray hotspot identified by the Telescope Array Collaboration (Abassi et al. 2014). If these magnetic field values are typical for this general region of the sky, the implication is that the highest-energy cosmic rays ($\gtrsim 6\times 10^{19} {\rm \, eV}$), among which the hotspot has been identified, get deflected by $\gtrsim 10^\circ$ over their propagation through the Galaxy {\em alone} if they are {\em protons} (see, e.g., Magkos \& Pavlidou 2018). This deflection magnitude is comparable to the radial extent of the excess as estimated by the Telescope Array Collaboration ($10^\circ$, Abassi et al. 2014). The implication in this case would be that the cosmic rays responsible for the Telescope Array hotspot {\em have to be protons}, as heavier nuclei would deflect more, proportionally to their atomic number Z, and hence produce a much more extended and less pronounced excess. However, the most recent Auger Collaboration results for the composition of ultra-high-energy cosmic rays (based on the reconstruction of air showers produced in collisions between cosmic rays and atmospheric atoms) detected by the southern-hemisphere, high-statistics cosmic-ray observatory favour a heavier composition, assuming the Standard Model of Particle Physics holds without modifications in ultra-LHC energies, up to 100 TeV. That our results (based only on electromagnetic propagation of cosmic rays in the Galactic field, i.e. physics much more certain to hold up to the highest energies) are not consistent with this finding can be explained in two ways: either the composition of the northern hotspot is different from that produced by the typical cosmic-ray source; or new physics sets in at center-of-mass energies $\sim 50$ TeV, as has been suggested for other reasons by several authors (e.g., Farrar \& Allen 2013; Anchordoqui et al. 2017; The Pierre Auger Collaboration et al. 2017a; Tomar 2017; Pavlidou \& Tomaras 2018).

\subsection{Future prospects}

In this paper we tomographically estimated the value of the POS magnetic field of only a very small region of the sky. However, available H\textsc{I} exist for the entire sky and up to very large distances ($\le$ 600 km~$\rm{s^{-1}}$) (HI4PI Collaboration et al. 2016). Such data can be processed with the Rolling Hough Transform (Clark et al. 2014) in velocity slices. Such an analysis was already performed for the entire second data release of the GALFA-H\textsc{I} survey, yielding the orientation angle of fibers. With the orientation angle known, the process of considering cuts perpendicular to fibers and applying the method developed by Tritsis et al. (2018) can be fully automated. Advancements in polarization measurements and the upcoming PASIPHAE (Polar-Areas Stellar-Imaging in Polarization High-Accuracy Experiment)\footnote{\url{http://pasiphae.science/}} optical-polarimetry survey (Tassis et al. 2018) will yield a 3D map of the POS orientation of the Galactic magnetic field. Thus, we aim to apply the method to large fractions of the sky in a follow-up study.

\section{Summary}\label{sum}

We used the theory of MHD waves and applied the method developed by Tritsis et al. (2018) to derive the POS component of the magnetic field from spectroscopic observations of fibers for two clouds along one line of sight close to Ursa Major. Our results were combined with existing measurements of the LOS magnetic field from Zeeman observations. We find that for both clouds the magnetic field is a factor of $\sim$ 5 larger than what theoretical models of the global Galactic magnetic field predict (Sun et al. 2008; Jansson \& Farrar 2012). More specifically, the median value of the magnetic field for the cloud further away from Earth is $15^{+8}_{-3}~\rm{\upmu G}$. For the cloud located closer to Earth the POS magnetic field ranges from $4^{+1}_{-1}$ to $17^{+14}_{-4	}~\rm{\upmu G}$ with the variations in the LOS component of the magnetic field in rough anti-correlation. 

The fact that the theoretical predictions from the model developed for striations (Tritsis \& Tassis 2016) applies in observations of fibers, strongly suggests that fibers are created from hydromagnetic waves. Finally, our results of the strength of the magnetic field have important implications about the nature of cosmic rays.

\section*{Acknowledgements}

We thank R. Skalidis, G. Panopoulou, K. Tassis, E. Ntormousi and G. Magkos for useful comments and discussions. We thank the anonymous referee for comments that helped improve this work. C.~F.~acknowledges funding provided by the Australian Research Council (Discovery Projects DP150104329 and DP170100603, and Future Fellowship FT180100495), and the Australia-Germany Joint Research Cooperation Scheme (UA-DAAD). 3D-Visualizations made with Space Nebula Plugin for Unreal Engine 4 (Fabian Fuchs \& Linus Fuchs, private communication: Thauros-Development@outlook.com). The data analysis presented in this work used high-performance computing resources provided by the Leibniz Rechenzentrum and the Gauss Centre for Supercomputing (grants~pr32lo, pr48pi and GCS Large-scale project~10391), the Partnership for Advanced Computing in Europe (PRACE grant pr89mu), the Australian National Computational Infrastructure (grant~ek9), and the Pawsey Supercomputing Centre with funding from the Australian Government and the Government of Western Australia, in the framework of the National Computational Merit Allocation Scheme and the ANU Allocation Scheme.

\appendix
\section{Distance estimates}\label{distance}
Recently, Wenger et al. (2018) developed a Monte-Carlo code for estimating the kinematic distances to clouds. Using their code and the derived velocity centres of the clouds, we find that the distance to Cloud 1 is 3.64 $\pm$ 0.96 kpc and the upper limit for the distance of Cloud 2 is 410 pc. However, kinematic distances are not so robust away from the Galactic disk. Green et al. (2018) provided a 3D map of interstellar dust reddening from Pan-STARRS 1 (Chambers et al. 2016) and 2MASS (Skrutskie et al. 2006) photometric data. Based on their map, in Figure~\ref{Red} we show the reddening as a function of distance for the coordinates where the two clouds are located. From the points where the reddening curves exhibit an abrupt increase it can be seen that the distance to Cloud 1 appears to be close to $\sim$ 0.8 -- 1 kpc (i.e. a factor of 3 less than its kinematic distance). The distance to Cloud 2 can be identified at $\sim$ 200 pc. Polarization measurements with the RoboPol instrument (King et al. 2014) at the Skinakas Observatory in Crete also place the first cloud between 200 and 500 pc (R. Skalidis - private communication). These distance estimates are used to crudely estimate the number density of each cloud (see \S~\ref{Bfield}), compare are results to global Galactic magnetic field models and in order to put our results in the greater context of cosmic-ray propagation implications. Here, we adopt a value of 1 kpc for the distance to Cloud 1 and 200 pc for the distance to Cloud 2. 

\setcounter{figure}{0}
\renewcommand{\thefigure}{A\arabic{figure}}

\begin{figure}[!thbp]
\centering
\includegraphics[width=0.7\columnwidth, clip]{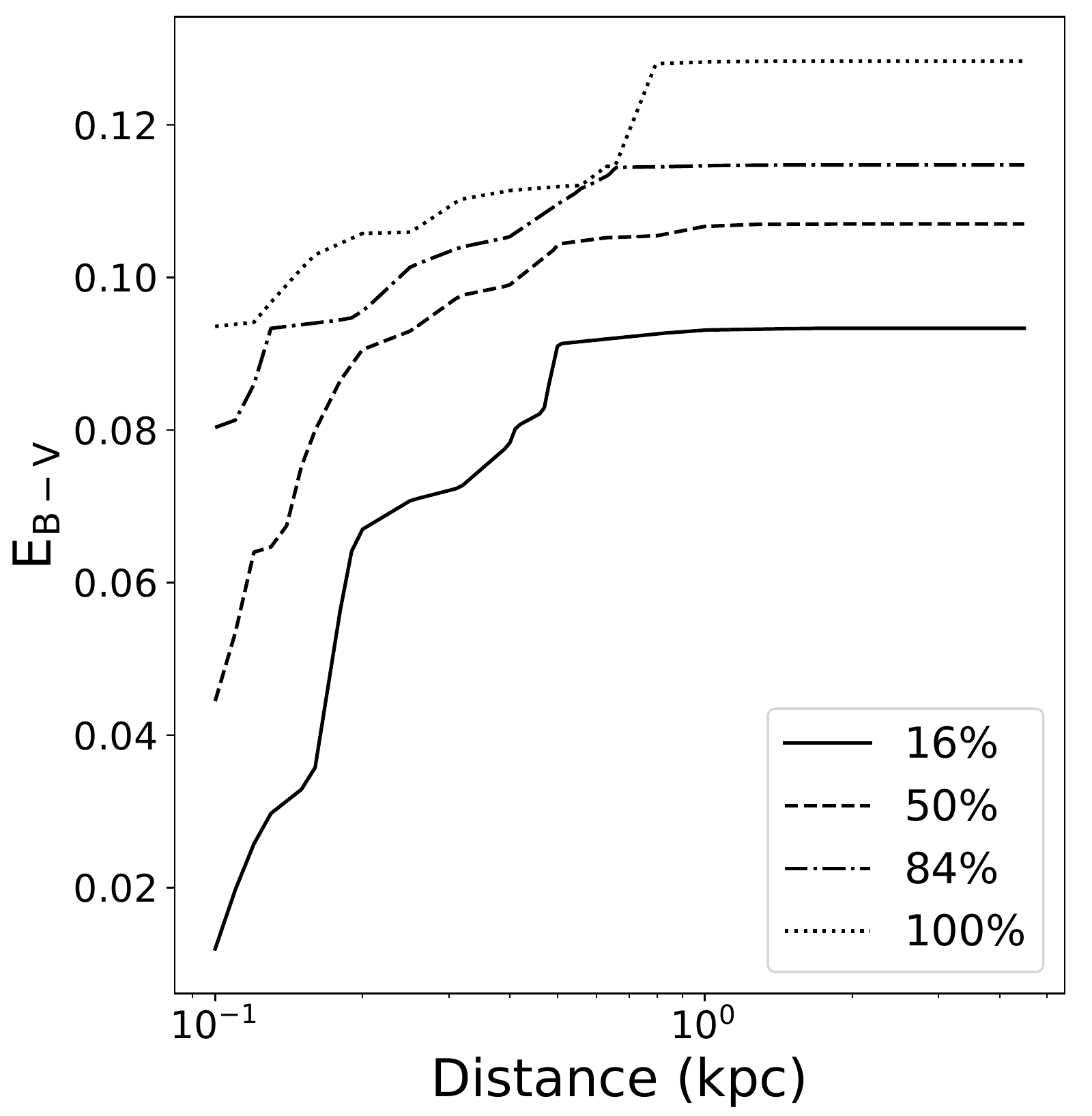}
\caption{Reddening as a function of distance from the dust reddening map by Green et al. (2018). Different lines represent different percentiles of reddening. The distance to Cloud 1 is estimated to $\sim$ 0.8 - 1 kpc (second sudden rise in reddening), less than computed kinematically and the distance to Cloud 2 is $\sim$ 200 - 500 pc (first sudden rise in reddening).
\label{Red}}
\end{figure}

\section{First-moment maps}\label{fmm}

In the upper and lower panels of Figure~\ref{fmm} we show the first moment maps of the velocity of Cloud 1 and Cloud 2 respectively.

\setcounter{figure}{0}
\renewcommand{\thefigure}{B\arabic{figure}}

\begin{figure}[!thbp]
\centering
\includegraphics[width=1.0\columnwidth, clip]{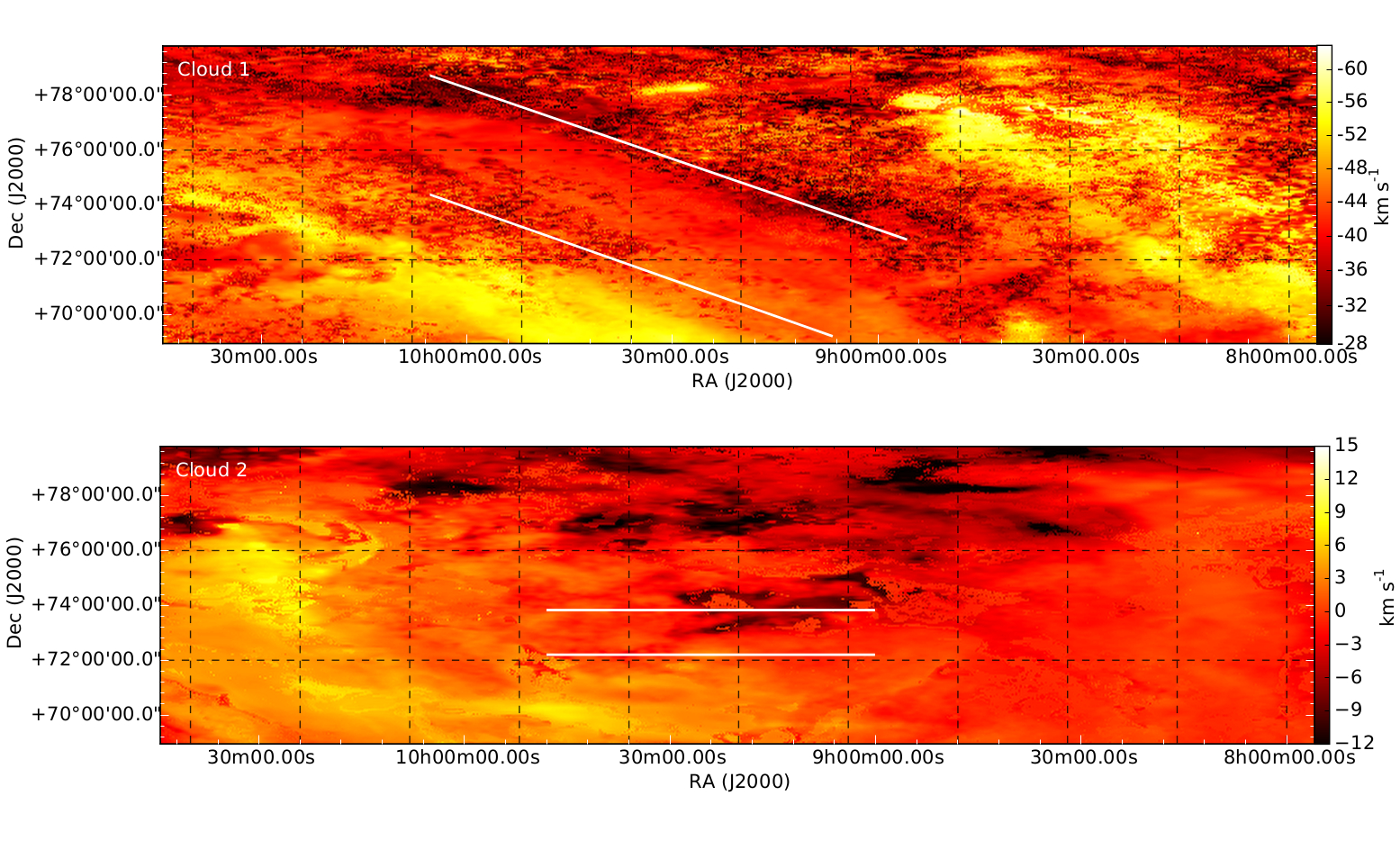}
\caption{H\textsc{I} velocity centroid maps of Cloud 1 (upper panel) and Cloud 2 (lower panel). The white lines in the two panels are the same as in Figure~\ref{Combo}.
\label{fmm}}
\end{figure}

\section{Error Estimation}\label{errors}

Errors in computing the spatial frequencies and the power in the power spectra of either the column density or the first moment of velocity have a twofold origin. The first is the noise level in the observations. The second origin is the fact that the endpoints of the cuts perpendicular to fibers	are not symmetric. In order to quantify the error in the column density power spectra we first consider a cut perpendicular to fibers, compute its power spectrum and find the peaks in that power spectrum. We then bootstrap one hundred times by adding noise to the cut, drawn from a Gaussian distribution, with $\sigma$ being the noise level in column density, computed from the noise level in the observations ($\sim$ 100 mK, Winkel et al. 2016) and Equation~\ref{cdDeriv}. At the same time, we vary the length of the cut (up to 80\% of its original length) in order to investigate how the asymmetry in the endpoints affects the derived frequencies and powers. We then compute the standard deviation of the frequencies and powers for each peak. We repeat the same process for the power spectra of velocities where, instead of an error from the noise level of the data, we use the mean error from the fits of Gaussians to the spectral lines. We use these errors to compute the uncertainties in the derived quantities via error propagation.


\begin{thebibliography}{99}

\bibitem[Abbasi et al.(2014)]{2014ApJ...790L..21A} Abbasi, R.~U., Abe, M., Abu-Zayyad, T., et al.\ 2014, \apjl, 790, L21 
\bibitem[Abbasi et al.(2018)]{2018arXiv180205003A} Abbasi, R.~U., Abe, M., Abu-Zayyad, T., et al.\ 2018, arXiv:1802.05003 
%\bibitem[Ackermann et al.(2012)]{2012ApJ...750....3A} Ackermann, M., Ajello, M., Atwood, W.~B., et al.\ 2012, \apj, 750, 3 
\bibitem[Alves de Oliveira et al.(2014)]{2014A&A...568A..98A} Alves de Oliveira, C., Schneider, N., Mer{\'{\i}}n, B., et al.\ 2014, \aap, 568, A98 
\bibitem[Anchordoqui et al.(2017)]{2017PhRvD..95f3005A} Anchordoqui, L.~A., Goldberg, H., \& Weiler, T.~J.\ 2017, \prd, 95, 063005 
\bibitem[Andersson et al.(2015)]{2015ARA&A..53..501A} Andersson, B.-G., Lazarian, A., \& Vaillancourt, J.~E.\ 2015, \araa, 53, 501 
\bibitem[BICEP2/Keck Collaboration et al.(2015)]{2015PhRvL.114j1301B} BICEP2/Keck Collaboration, Planck Collaboration, Ade, P.~A.~R., et al.\ 2015, Physical Review Letters, 114, 101301 
\bibitem[Burgers (1948)]{1948ADVAPPLMECH..1..171} Burgers, J.-M.\ 1948, Advances in Applied Mechanics, 1, 171 
\bibitem[Caldwell et al.(2017)]{2017ApJ...839...91C} Caldwell, R.~R., Hirata, C., \& Kamionkowski, M.\ 2017, \apj, 839, 91 
\bibitem[Chepurnov et al.(2010)]{2010ApJ...714.1398C} Chepurnov, A., Lazarian, A., Stanimirovi{\'c}, S., Heiles, C., \& Peek, J.~E.~G.\ 2010, \apj, 714, 1398 
\bibitem[Chambers et al.(2016)]{2016arXiv161205560C} Chambers, K.~C., Magnier, E.~A., Metcalfe, N., et al.\ 2016, arXiv:1612.05560 
\bibitem[Chandrasekhar \& Fermi(1953)]{1953ApJ...118..113C} Chandrasekhar, S., \& Fermi, E.\ 1953, \apj, 118, 113
\bibitem[Chengalur et al.(2013)]{2013MNRAS.432.3074C} Chengalur, J.~N., Kanekar, N., \& Roy, N.\ 2013, \mnras, 432, 3074 
\bibitem[Clark et al.(2014)]{2014ApJ...789...82C} Clark, S.~E., Peek, J.~E.~G., \& Putman, M.~E.\ 2014, \apj, 789, 82 
\bibitem[Clark et al.(2015)]{2015PhRvL.115x1302C} Clark, S.~E., Hill, J.~C., Peek, J.~E.~G., Putman, M.~E., \& Babler, B.~L.\ 2015, Physical Review Letters, 115, 241302 
\bibitem[Clark(2018)]{2018ApJ...857L..10C} Clark, S.~E.\ 2018, \apjl, 857, L10 
\bibitem[Cox et al.(2016)]{2016A&A...590A.110C} Cox, N.~L.~J., Arzoumanian, D., Andr{\'e}, P., et al.\ 2016, \aap, 590, A110
\bibitem[Davis (1951)]{1951Phys.Rev....81...890} Davis, L.\ 1951, Phys. Rev., 81, 890
\bibitem[Dickey \& Lockman(1990)]{1990ARA&A..28..215D} Dickey, J.~M., \& Lockman, F.~J.\ 1990, \araa, 28, 215 
\bibitem[Farrar \& Allen(2013)]{2013EPJWC..5307007F} Farrar, G.~R., \& Allen, J.~D.\ 2013, European Physical Journal Web of Conferences, 53, 07007
\bibitem[Federrath \& Klessen(2013)]{2013ApJ...763...51F} Federrath, C., \& Klessen, R.~S.\ 2013, \apj, 763, 51
\bibitem[Gazol \& Villagran(2018)]{2018MNRAS.478..146G} Gazol, A., \& Villagran, M.~A.\ 2018, \mnras, 478, 146 
\bibitem[Goldsmith et al.(2008)]{2008ApJ...680..428G} Goldsmith, P.~F., Heyer, M., Narayanan, G., et al.\ 2008, \apj, 680, 428-445 
\bibitem[Gonz{\'a}lez-Casanova \& Lazarian(2017)]{2017ApJ...835...41G} Gonz{\'a}lez-Casanova, D.~F., \& Lazarian, A.\ 2017, \apj, 835, 41 
\bibitem[Goodman et al.(1994)]{1994ASPC...58..425G} Goodman, A.~A., Myers, P.~C., Gusten, R., \& Heiles, C.\ 1994, The First Symposium on the Infrared Cirrus and Diffuse Interstellar Clouds, 58, 425
\bibitem[Green et al.(2018)]{2018MNRAS.478..651G} Green, G.~M., Schlafly, E.~F., Finkbeiner, D., et al.\ 2018, \mnras, 478, 651 
\bibitem[Haverkorn(2015)]{2015ASSL..407..483H} Haverkorn, M.\ 2015, Magnetic Fields in Diffuse Media, 407, 483 
\bibitem[Heiles(1989)]{1989ApJ...336..808H} Heiles, C.\ 1989, \apj, 336, 808 
\bibitem[Heiles \& Troland(2003)]{2003ApJ...586.1067H} Heiles, C., \& Troland, T.~H.\ 2003, \apj, 586, 1067 
\bibitem[Heiles \& Troland(2004)]{2004ApJS..151..271H} Heiles, C., \& Troland, T.~H.\ 2004, \apjs, 151, 271 
\bibitem[HI4PI Collaboration et al.(2016)]{2016A&A...594A.116H} HI4PI Collaboration, Ben Bekhti, N., Fl{\"o}er, L., et al.\ 2016, \aap, 594, A116 
\bibitem[Inoue \& Inutsuka(2016)]{2016ApJ...833...10I} Inoue, T., \& Inutsuka, S.-i.\ 2016, \apj, 833, 10 
%\bibitem[Hunter et al.(1997)]{1997ApJ...481..205H} Hunter, S.~D., Bertsch, D.~L., Catelli, J.~R., et al.\ 1997, \apj, 481, 205 
\bibitem[Jansson \& Farrar(2012)]{2012ApJ...761L..11J} Jansson, R., \& Farrar, G.~R.\ 2012, \apjl, 761, L11 
\bibitem[Jeli{\'c} et al.(2018)]{2018arXiv180606634J} Jeli{\'c}, V., Prelogovi{\'c}, D., Haverkorn, M., Remeijn, J., \& Klind{\v z}i{\'c}, D.\ 2018, arXiv:1806.06634 
\bibitem[Kalberla et al.(2010)]{2010A&A...521A..17K} Kalberla, P.~M.~W., McClure-Griffiths, N.~M., Pisano, D.~J., et al.\ 2010, \aap, 521, A17 
\bibitem[Kalberla \& Haud(2015)]{2015A&A...578A..78K} Kalberla, P.~M.~W., \& Haud, U.\ 2015, \aap, 578, A78 
\bibitem[Kerp et al.(2011)]{2011AN....332..637K} Kerp, J., Winkel, B., Ben Bekhti, N., Fl{\"o}er, L., \& Kalberla, P.~M.~W.\ 2011, Astronomische Nachrichten, 332, 637
\bibitem[King et al.(2014)]{2014MNRAS.442.1706K} King, O.~G., Blinov, D., Ramaprakash, A.~N., et al.\ 2014, \mnras, 442, 1706  
\bibitem[Magkos \& Pavlidou(2018)]{2018arXiv180203409M} Magkos, G., \& Pavlidou, V.\ 2018, arXiv:1802.03409 
\bibitem[Magnani et al.(1996)]{1996ApJS..106..447M} Magnani, L., Hartmann, D., \& Speck, B.~G.\ 1996, \apjs, 106, 447 
\bibitem[Malinen et al.(2016)]{2016MNRAS.460.1934M} Malinen, J., Montier, L., Montillaud, J., et al.\ 2016, \mnras, 460, 1934
\bibitem[McClure-Griffiths et al.(2006)]{2006ApJ...652.1339M} McClure-Griffiths, N.~M., Dickey, J.~M., Gaensler, B.~M., Green, A.~J., \& Haverkorn, M.\ 2006, \apj, 652, 1339 
\bibitem[McClure-Griffiths et al.(2009)]{2009ApJS..181..398M} McClure-Griffiths, N.~M., Pisano, D.~J., Calabretta, M.~R., et al.\ 2009, \apjs, 181, 398 
\bibitem[Miville-Desch{\^e}nes et al.(2010)]{2010A&A...518L.104M} Miville-Desch{\^e}nes, M.-A., Martin, P.~G., Abergel, A., et al.\ 2010, \aap, 518, L104 
\bibitem[Myers et al.(1995)]{1995ApJ...442..177M} Myers, P.~C., Goodman, A.~A., Gusten, R., \& Heiles, C.\ 1995, \apj, 442, 177 
\bibitem[Orlando \& Strong(2013)]{2013MNRAS.436.2127O} Orlando, E., \& Strong, A.\ 2013, \mnras, 436, 2127 
\bibitem[Palmeirim et al.(2013)]{2013A&A...550A..38P} Palmeirim, P., Andr{\'e}, P., Kirk, J., et al.\ 2013, \aap, 550, A38 
\bibitem[Panopoulou et al.(2016)]{2016MNRAS.462.1517P} Panopoulou, G.~V., Psaradaki, I., \& Tassis, K.\ 2016, \mnras, 462, 1517 
\bibitem[Pavlidou \& Tomaras(2018)]{2018arXiv180204806P} Pavlidou, V., \& Tomaras, T.\ 2018, arXiv:1802.04806 
\bibitem[Peek et al.(2011)]{2011ApJS..194...20P} Peek, J.~E.~G., Heiles, C., Douglas, K.~A., et al.\ 2011, \apjs, 194, 20 
\bibitem[Peek et al.(2018)]{2018ApJS..234....2P} Peek, J.~E.~G., Babler, B.~L., Zheng, Y., et al.\ 2018, \apjs, 234, 2 
\bibitem[The Pierre Auger Collaboration et al.(2017a)]{2017arXiv170806592T} The Pierre Auger Collaboration, Aab, A., Abreu, P., et al.\ 2017a, arXiv:1708.06592 
\bibitem[Pierre Auger Collaboration et al.(2017b)]{2017Sci...357.1266P} Pierre Auger Collaboration, Aab, A., Abreu, P., et al.\ 2017b, Science, 357, 1266 
\bibitem[Planck Collaboration et al.(2011)]{2011A&A...536A..24P} Planck Collaboration, Abergel, A., Ade, P.~A.~R., et al.\ 2011, \aap, 536, A24 
\bibitem[Planck Collaboration et al.(2016)]{2016A&A...586A.135P} Planck Collaboration, Adam, R., Ade, P.~A.~R., et al.\ 2016a, \aap, 586, A135 
\bibitem[Planck Collaboration et al.(2016)]{2016A&A...596A.103P} Planck Collaboration, Adam, R., Ade, P.~A.~R., et al.\ 2016b, \aap, 596, A103 
\bibitem[Planck Collaboration et al.(2016)]{2016A&A...586A.133P} Planck Collaboration, Adam, R., Ade, P.~A.~R., et al.\ 2016c, \aap, 586, A133 
\bibitem[Planck Collaboration et al.(2016)]{2016A&A...596A.105P} Planck Collaboration, Aghanim, N., Alves, M.~I.~R., et al.\ 2016d, \aap, 596, A105 
\bibitem[Planck Collaboration et al.(2018)]{2018arXiv180706212P} Planck Collaboration, Aghanim, N., Akrami, Y., et al.\ 2018, arXiv:1807.06212 
\bibitem[Robishaw(2008)]{2008PhDT........13R} Robishaw, T.\ 2008, \textit{Magnetic fields near and far: galactic and extragalactic single-dish radio observations of the Zeeman effect.} Ph.D.~Thesis, Univ. Calif., Berkeley (AAT 3331778) 
\bibitem[Solomon et al.(1987)]{1987ApJ...319..730S} Solomon, P.~M., Rivolo, A.~R., Barrett, J., \& Yahil, A.\ 1987, \apj, 319, 730 
\bibitem[Skrutskie et al.(2006)]{2006AJ....131.1163S} Skrutskie, M.~F., Cutri, R.~M., Stiening, R., et al.\ 2006, \aj, 131, 1163 
\bibitem[Spruit(2013)]{2013arXiv1301.5572S} Spruit, H.~C.\ 2013, arXiv:1301.5572 
\bibitem[Sun et al.(2008)]{2008A&A...477..573S} Sun, X.~H., Reich, W., Waelkens, A., \& En{\ss}lin, T.~A.\ 2008, \aap, 477, 573
\bibitem[Tassis \& Pavlidou(2015)]{2015MNRAS.451L..90T} Tassis, K., \& Pavlidou, V.\ 2015, \mnras, 451, L90
\bibitem[Tassis et al.(2018)]{2018arXiv181005652T} Tassis, K., Ramaprakash, A.~N., Readhead, A.~C.~S., et al.\ 2018, arXiv:1810.05652 
\bibitem[Tomar(2017)]{2017PhRvD..95i5035T} Tomar, G.\ 2017, \prd, 95, 095035  
\bibitem[Tritsis \& Tassis(2016)]{2016MNRAS.462.3602T} Tritsis, A., \& Tassis, K.\ 2016, \mnras, 462, 3602
\bibitem[Tritsis \& Tassis(2018)]{2018Sci...360..635T} Tritsis, A., \& Tassis, K.\ 2018, Science, 360, 635 
\bibitem[Tritsis et al.(2018)]{2018MNRAS.481.5275T} Tritsis, A., Federrath, C., Schneider, N., \& Tassis, K.\ 2018, \mnras, 481, 5275 
\bibitem[Wenger et al.(2018)]{2018ApJ...856...52W} Wenger, T.~V., Balser, D.~S., Anderson, L.~D., \& Bania, T.~M.\ 2018, \apj, 856, 52 
\bibitem[Winkel et al.(2016)]{2016A&A...585A..41W} Winkel, B., Kerp, J., Fl{\"o}er, L., et al.\ 2016, \aap, 585, A41 

%==============================================================%


%\bibitem[Crutcher(2012)]{2012mfu3.conf...31C} Crutcher, R.~M.\ 2012, Magnetic Fields in the Universe III - From Laboratory and Stars to Primordial Structures, proceedings of the conference held 21-27 August, 2011 in Zakopane, Poland.~Edited by M.~Soida et al.~Jagiellonian University, Astronomical Observatory, 2012, p.31, 31 
%\bibitem[Crutcher(2012)]{2012ARA&A..50...29C} Crutcher, R.~M.\ 2012, \araa, 50, 29
 
%\bibitem[Falgarone et al.(2009)]{2009A&A...507..355F} Falgarone, E., Pety, J., \& Hily-Blant, P.\ 2009, \aap, 507, 355 
%\bibitem[Griffin et al.(2010)]{2010A&A...518L...3G} Griffin, M.~J., Abergel, A., Abreu, A., et al.\ 2010, \aap, 518, L3 
%\bibitem[McElroy et al.(2013)]{2013A&A...550A..36M} McElroy, D., Walsh, C., Markwick, A.~J., et al.\ 2013, \aap, 550, A36 
%\bibitem[Ossenkopf \& Mac Low(2002)]{2002A&A...390..307O} Ossenkopf, V., \& Mac Low, M.-M.\ 2002, \aap, 390, 307 
%\bibitem[Pearson et al.(2014)]{2014ExA....37..175P} Pearson, C., Lim, T., North, C., et al.\ 2014, Experimental Astronomy, 37, 175 
%\bibitem[Roman-Duval et al.(2011)]{2011ApJ...740..120R} Roman-Duval, J., Federrath, C., Brunt, C., et al.\ 2011, \apj, 740, 120 
%\bibitem[Roy et al.(2013)]{2013ApJ...763...55R} Roy, A., Martin, P.~G., Polychroni, D., et al.\ 2013, \apj, 763, 55 
%\bibitem[Schlafly et al.(2014)]{2014ApJ...786...29S} Schlafly, E.~F., Green, G., Finkbeiner, D.~P., et al.\ 2014, \apj, 786, 29 


%==============================================================% 
\end{thebibliography}
\end{document}